\documentclass[twoside]{article}
\usepackage{fleqn,espcrc2}

\usepackage{graphicx}

\def\brg{{B \rightarrow \rho \gamma}}
\def\brog{{B \rightarrow \rho (\omega) \gamma}}
\def\bkg{{B \rightarrow K^\ast \gamma}}
\def\bsg{{B \rightarrow X_s \gamma}}
\def\bdg{{B \rightarrow X_d \gamma}}
\def\nbrg{{B^0 \rightarrow \rho^0 \gamma}}
\def\cbrg{{B^\pm \rightarrow \rho^\pm \gamma}}

\def\acp{{\cal A}_{\rm CP}}
\def\b{{\cal B}}
\def\c{{\cal C}}
\def\o{{\cal O}}
\def\m{{\cal M}}
\def\he{{\cal H}_{\rm eff}}
\def\g{\Gamma}
\def\e{{\rm e}}
\def\l{\lambda}

\title{Isospin symmetry breaking and direct CP violation 
  in $\brg$ within and beyond the standard model 
  \thanks{Based on the work in \cite{ahl}.}}

\author{L.T. Handoko \address[desy]{
        Deutsches Elektronen Synchrotron, DESY, \\ 
        Notkestr. 85, 22609 Hamburg, Germany}}

\begin{document}

\begin{abstract}
We discuss the isospin symmetry breaking quantity 
$\Delta^{\pm 0} \equiv {\Gamma(\cbrg)}/{2 \Gamma(\nbrg)} - 1$, 
and the direct CP violation in $\cbrg$ within and 
beyond the standard model. After showing that the 
leading-order calculation would be a good approximation 
in such models, we argue that measurements of these 
quantities would be able to disentagle physics beyond 
the standard model.  
\vspace{1pc}
\end{abstract}

\maketitle

\section{INTRODUCTION}

It is highly expected that the radiative decays $\brog$ 
would provide independent information on the  
Cabibbo-Kobayashi-Maskawa (CKM) matrix, i.e. 
the parameters $\rho$ and $\eta$ in the Wolfenstein 
parametrization. This expectation is motivated 
by successful measurements of the radiative 
decays $\bkg$ \cite{bkg} and its inclusive mode $\bsg$ \cite{bsg}. 
A theoretically robust way to achieve this is by considering the 
ratio of $\bdg$ and $\bsg$ decay rates, which is essentially 
proportional to the CKM ratio $|{V_{td}}/{V_{ts}}|^2$. 
Hence once $\bdg$ is measured,   
this ratio would provide a measurement of $|{V_{td}}/{V_{ts}}|$ 
in an independent manner. We recall that the branching ratios 
for $\bsg$ \cite{ghw} and $\bdg$ \cite{bdgt} in the standard model 
(SM) are known in next-to-leading-order (NLO).

However, as it is a challenge to measure $\bdg$, experimentally 
the exclusive decays $\brog$ are more favored and would be 
available much earlier, as suggested by the present experimental 
limits on $\brog$ \cite{bkg}.  As pointed long time ago, 
in the charged decays $\cbrg$, the 
$W-$ annihilation \cite{ab} and also long-distance 
(LD) effects \cite{lde} could be significant and contribute 
$\sim 20\%$ to the total rate. Actually these contributions induce 
violation of isospin symmetry, defined as
\begin{equation}
  \Delta^{\pm 0} \equiv \frac{\Gamma(\cbrg)}{2 \Gamma(\nbrg)} 
  - 1 \; . 
  \label{eq:isb}
\end{equation}
Of great interest is also direct CP asymmetry, 
\begin{equation}
        \acp \equiv 
        \frac{\b( B^- \rightarrow \rho^- \gamma)
          - \b(B^+ \rightarrow \rho^+ \gamma)}{
          \b(B^- \rightarrow \rho^- \gamma)
          + \b(B^+ \rightarrow \rho^+ \gamma)} \; . 
  \label{eq:acp}
\end{equation}
We are going to show that the quantities 
$\Delta^{\pm 0}$ and $\acp$ are good probes 
to search for physics beyond the standard model. 

\section{THE RADIATIVE DECAYS $\brg$}

Now, let us concentrate on the radiative exclusive 
decays $\brg$. The processes are governed by the 
following amplitude
\begin{equation}
  \m = \left< \rho(p_\rho) \left| \he \right| B(p_B) \right> \; ,
   \label{eqn:m}
\end{equation}
where the effective Hamiltonian describes the radiative 
weak-transition $b \rightarrow d \gamma$ 
\begin{eqnarray}
\he & = & \frac{G_F}{\sqrt{2}}\left[ 
  \lambda_{u}^{(d)}(\c_1(\mu) {\cal O}_1(\mu) 
  + \c_2(\mu) {\cal O}_2(\mu)) 
  \right. \nonumber \\
& &  \left. - \lambda_{d}^{(t)} \c_7^{eff}(\mu)
   {\cal O}_7(\mu) +...\right].
\end{eqnarray}
Here, $\lambda_{q}^{(q^\prime)} = V_{qb}V_{qq^\prime}^*$ are the CKM
factors, and we have restricted ourselves to those contributions which
will be important in what follows. The operators ${\cal O}_1(\mu)$ and
${\cal O}_2(\mu)$ are the four-quark operators
\begin{eqnarray}
{\cal O}_1 & = & (\bar{d}_\alpha \Gamma^\mu u_\beta)(\bar{u}_\beta
\Gamma_\mu b_\alpha) \; , \\
{\cal O}_2 & = & (\bar{d}_\alpha \Gamma^\mu u_\alpha)(\bar{u}_\beta 
\Gamma_\mu b_\beta) \; , \\
{\cal O}_7 & = & \frac{e m_b}{8 \pi^2} \bar{d} \sigma^{\mu
\nu}(1-\gamma_5) F_{\mu \nu} b \; , 
\end{eqnarray}
where $\Gamma_\mu=\gamma_\mu(1-\gamma_5)$, $\alpha$ and $\beta$ are
the SU(3) color indices, $\c_1$ and $\c_2$ are the corresponding
Wilson coefficients, and 
$F_{\mu \nu}$ is the electromagnetic field strength tensor.

Using the parametrisation of the form-factor, one obtains 
the decay width for the charged and neutral decays as 
\begin{eqnarray}
\g^\pm & = & 
  \frac{G_F^2 \alpha |\l^{(d)}_t|^2}{32 \pi^4} m_b^2 m_B^3
  \left( 1 - \frac{m_\rho^2}{m_B^2} \right)^3 
  \left| T_1^\rho \right|^2 
  \nonumber \\
  & & \times \left| 1 -  
    \frac{\vert \lambda_{u}^{(d)}\vert}{\vert \lambda_{t}^{(d)}\vert}
    \epsilon_A \e^{i (\phi_A \mp \alpha)} \right|^2 \; , 
  \label{eq:gpm} \\
\g^0 & = & 
  \frac{G_F^2 \alpha |\l^{(d)}_t|^2}{32 \pi^4} m_b^2 m_B^3
  \left( 1 - \frac{m_\rho^2}{m_B^2} \right)^3 
  \left| T_1^\rho \right|^2 \; .
  \label{eq:g0}
\end{eqnarray}
Here, $T_1^\rho$ is the form-factor for the magnetic moment 
operator ($\c_7$) in the $B \rightarrow \rho$ transition 
with an on-shell photon emission. On the other hand, the 
second term in Eq. (\ref{eq:gpm}) denotes the dominant 
$W-$annihilation and the possible sub-dominant LD contributions.
Keeping only the dominant $W-$annihilation, it reads \cite{ab}
\begin{equation}
       \epsilon_A \e^{i \phi_A} \equiv
        \frac{4 \pi^2 m_\rho}{m_b}
      \frac{\c_2 + {\c_1}/{N_c}}{\c_7^{eff}} r_u^\rho \; ,
      \label{eq:epsa}
\end{equation}
where $r_u^\rho$ is the ratio of the SD and 
LD form-factors induced from the penguin and the $W-$annihilation diagrams 
\cite{ab} respectively, while $\phi_A$ parametrises the possible strong 
phase induced there. 

In fact, the effective Wilson coefficient $\c_7^{eff}$ and 
the matrix elements (vertex and gluon bremstrahlung) have been 
calculated up to NLO \cite{ghw}. 
Also taking into account the relation between $b$ quark 
pole mass and $\overline{MS}$ mass up to NLO accuracy, 
Eq. (\ref{eq:gpm}) becomes
\begin{eqnarray}
  \Gamma^\pm & = & \frac{G_F^2 \alpha |\l_t^{(d)}|^2}{32 \pi^4} m_B^5 
  \left( 1 - \frac{m_\rho^2}{m_B^2} \right)^3 
  \left| T_1^V \right|^2  \nonumber \\
    & & \times
    \left\{ 
    \left| \c_7^{(0)eff} + A_R^{(1)t} \right|^2  
    \right. \nonumber \\
    & & \left.
    + \left( F_1^2 + F_2^2 \right) \left( 
      \left| A_R^u + L_R^u \right|^2 \right)
    \right. \nonumber \\
    & & \left.
    + 2 F_1 \left[ \c_7^{(0)eff} 
      \left( A_R^u + L_R^u \right) + A_R^{(1)t} L_R^u \right] 
    \right. \nonumber \\
    & & \left.
    \mp 2 F_2 \left[\c_7^{(0)eff}  A_I^u 
     - A_I^{(1)t} L_R^u   \right] 
  \right\} \; ,
  \label{eq:gpmnlo}
\end{eqnarray}
where 
$F_1 = -|{\l_u^{(d)}}/{\l_t^{(d)}}| \cos\alpha$ 
and $F_2= -|{\l_u^{(d)}}/{\l_t^{(d)}}| \sin\alpha$, 
and $\alpha$ is one of the angles of the CKM unitarity triangle.
The analogue expression for Eq. (\ref{eq:g0}) can be obtained 
by putting $F_1=F_2=0$. 
While $L_R^u=\epsilon_A \c_7^{(0)eff}$,  $A_{R,I}^{(1)t}$ and
$A_{R,I}^u$ are functions of the real and imaginary parts of 
effective Wilson coefficients and
explicit $O(\alpha_s)$ contributions to the matrix elements 
evaluated at a scale $\mu$, 
\begin{eqnarray}
   A^{(1)t} & = & \frac{\alpha_s(m_b)}{4\pi} \left\{ 
   C_7^{(1)}(\mu) - \frac{16}{3} \, C_7^{(0)eff}(\mu) 
   \right. \\
   & & \left. 
     + \sum_i^8 C_i^{(0)eff}(\mu) \, \left[
       \gamma_{i7}^{(0)} \, \ln \frac{m_b}{\mu} + r_i(z) \right] 
     \right\} \; , \nonumber \\
   A^{(1)u} & = & \frac{\alpha_s(m_b)}{4\pi} \,
     C_2^{(0)}(\mu) \, \left[ r_2(z) - r_2(0) \right] \; , 
\end{eqnarray}
where $r_i$'s are complex numbers and 
$z = ({m_c}/{m_b})^2$. 

\section{ISOSPIN SYMMETRY BREAKING AND DIRECT CP VIOLATION}

According to the definition given in Eq. (\ref{eq:isb}), 
it might be better to further define the charged-conjugated 
averaged ratio as 
$\Delta \equiv \frac{1}{2} \left[ \Delta^{-0} + \Delta^{+0} \right]$
that would experimentally be more accessible. It has the
following expression, 
\begin{eqnarray}
  \Delta & = & 
  2 \epsilon_A \left\{ 
    F_1 +{1 \over 2}\epsilon_A(F_1^2 +F_2^2)
  \right. \nonumber \\
  & & \left. - \frac{1}{\c_7^{(0)eff}} \left[
    F_1 A_R^{(1)t} - \left( F_2^2 - F_1^2 \right) A_R^u 
    \right. 
  \right. \nonumber \\
  & & \left. \left. 
    + \epsilon_A \left( F_1^2 + F_2^2 \right) \left(A_R^{(1)t} + F_1 A_R^u \right)
    \right] \right\} \; .
\label{eq:delta} 
\end{eqnarray}
The first line is the leading-order (LO) expression, while 
the rest after it is the NLO corrections. 

For the CP asymmetry, using Eqs. (\ref{eq:acp}) and 
(\ref{eq:gpmnlo}), one obtains 
\begin{equation}
  \acp = -\frac{2 F_2}{\c_7^{(0)eff}( 1 + \Delta_{\rm LO} )} 
  \left[ A_I^u  - \epsilon_A A_I^t  \right] \; ,
\label{eq:acpnlo}
\end{equation}
where $\Delta_{\rm LO}$ is the LO part in Eq. (\ref{eq:delta}).
Of course, there would be additional source for the CP 
asymmetry if the strong phase $\phi_A$ is non-zero. Conversely, 
as long as $\phi_A = 0$, the CP asymmetry arises first 
at NLO where the strong phase is generated by higher order 
perturbative corrections.

According to Eqs. (\ref{eq:gpmnlo}) and (\ref{eq:acpnlo}), 
$\Delta$ is essentially proportional  
to $F_1$, while $\acp$ is proportional to $F_2$. 
Therefore, these quantities 
would provide complementary measurements to determine 
the angle $\alpha$. 

\section{RESULTS AND CONCLUSION}

\begin{figure}[t!]
\includegraphics[scale=0.4]{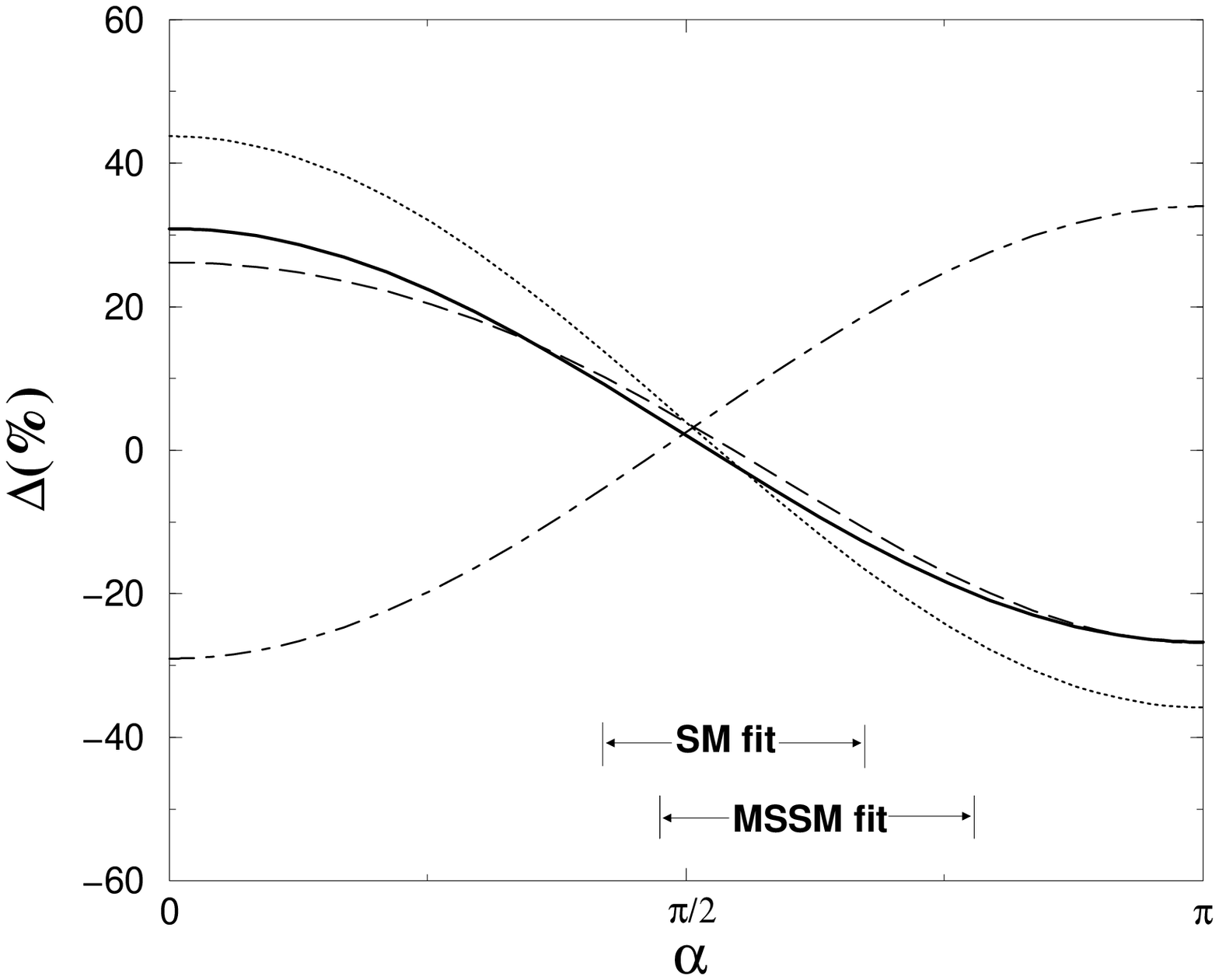}
\includegraphics[scale=0.4]{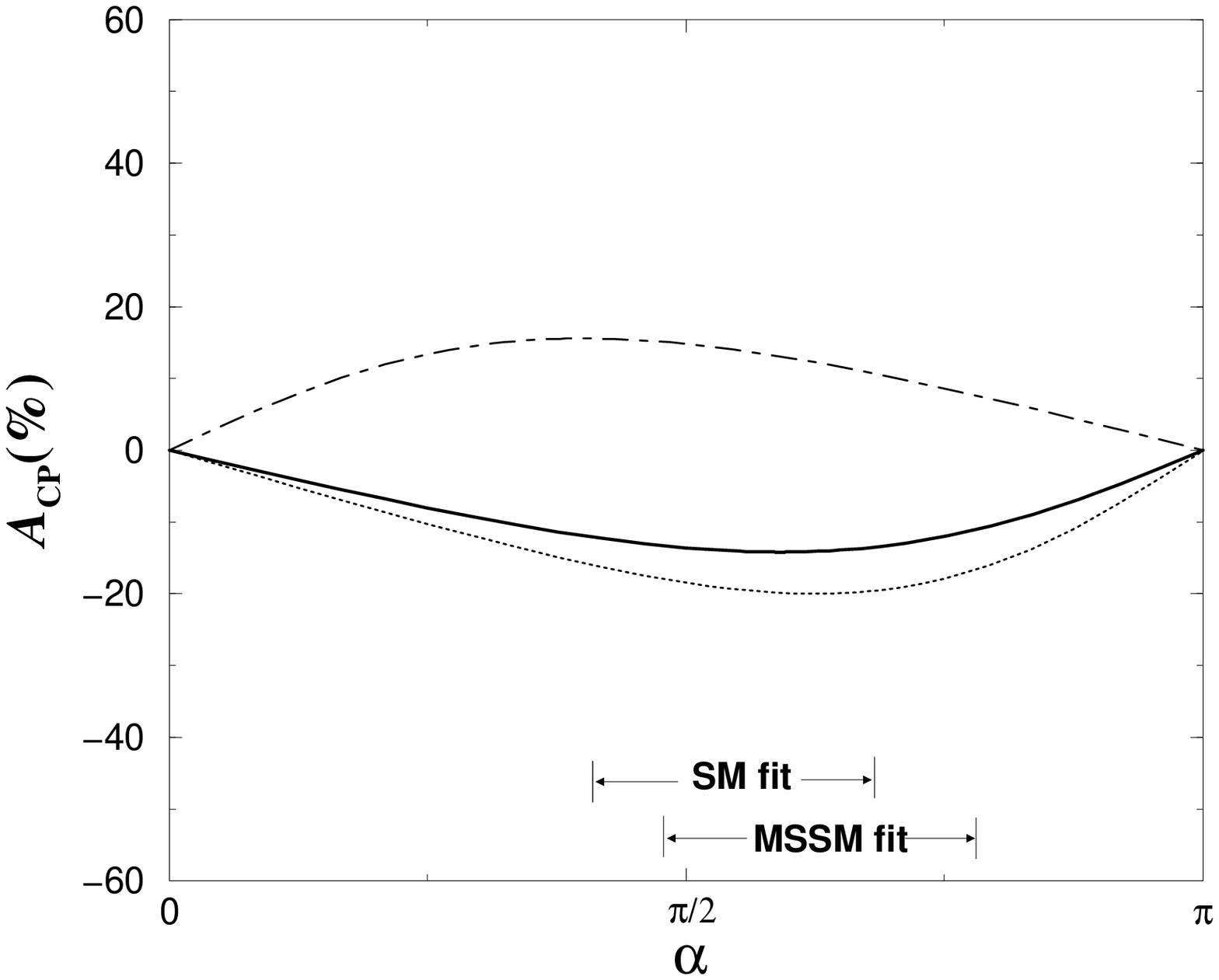}
\caption{$\Delta$ (upper) and $\acp$ (lower) in the 
  SM with (solid lines) and without (dashed lines) 
  NLO corrections, and in the MSSM with large 
  (dot-dashed lines) and small (dotted lines) 
  $\tan \beta$.}
\label {fig}
\end{figure}

First of all, let us stress that 
the function $A^{(1)u}$ is universal for all models which 
forbid the tree-level flavor-changing-neutral-current 
interaction, as $A^{(1)u}$ is generated by the 
$u \bar{u}$ loop through the charge current interactions 
denoted by the operators $\o_1$ and $\o_2$. 
This criterion is satisfied by the SM and in most variants 
of the supersymmetric (SUSY) model. Since we are going to 
consider only such models which satisfy this criterion, 
we can take the same values as the SM for 
$A_I^t$ and $A_I^u$, i.e. 
$A_{I}^{t}=-0.016$ and $A_{I}^{u}=+0.046$.
 
For the SUSY model, we take the minimal SUSY standard model 
(MSSM) as a particular example. This model is appropriate 
to show two extreme cases where the magnitude of $\c_7^{(0)eff}$ 
could be close to the SM prediction, but its sign is either the same 
(in the small $\tan \beta$ region), or opposite 
(in the large $\tan \beta$ region) as $\c_7^{(0)eff\rm (SM)}$. 
In the numerical analysis, we use the following values 
for the parameters :  
$(\tan \beta,{\c_7^{(0)eff}}/{\c_7^{(0)eff\rm (SM)}}) = (3,0.95)$ 
for small $\tan \beta$ region and 
$(\tan \beta, {\c_7^{(0)eff}}/{\c_7^{(0)eff\rm (SM)}}) = (30,-1.2)$ 
for large $\tan \beta$ region \cite{goto99}. The central value for 
the magnitude of CKM ratio in the SM
is $|{V_{ub}}/{V_{td}}| = 0.48$, while  in the MSSM  
$|{V_{ub}}/{V_{td}}| = 0.63$ which corresponds to $f = 0.6$ in \cite{al99} 
represents the maximum allowed contributions to $\Delta M_{d,s}$ 
and $|\epsilon_K|$, with 
$\Delta M_{d,s} = \Delta M_{d,s}^{\rm SM} (1 + f)$. 
Adopting the common 
knowledge for the $W-$annihilation, we take $\epsilon_A = -0.3$. 

The results are shown in Fig. (\ref{fig}). The SM and MSSM fits 
represent the allowed range of $\alpha$ (at 95\% C.L.) from fits 
to the unitarity triangle in each model \cite{al99}. We should emphasize 
that, by definition, the angle $\alpha$ and $F_{1,2}$ are 
correlated with each other and one should keep track the uncertainties 
in the $\alpha-F_{1,2}$ correlation. Details can be found 
in the original work \cite{ahl}.

From the figures, it is clear that the NLO contributions 
in $\Delta$ are 
small, and then one can take into account the LO contributions 
as a good approximation. Although the non-zero CP asymmetry is 
induced at NLO accuracy, it requires only the LO $\c_7^{(0)eff}$ 
as shown explicitely in Eq. (\ref{eq:acpnlo}). While the NLO 
corrections would enter through the sub-leading terms which 
are suppressed by $\epsilon_A$. 

Since both quantities are essentially proportional to the 
inverse of $\c_7^{(0)eff}$, they are sensitive to the sign 
of $\c_7^{(0)eff}$. For instance, if the large $\tan \beta$ 
MSSM solution is realized in nature, then the measured values 
of $\Delta$ and $\acp$ could be markedly different than in 
the SM. These would be striking signatures of new physics 
and strongly suggest the presence of supersymmetry. 

It should be stressed here that  
we have ignored all contributions where the photon and 
gluon lines are attached to the spectator line in 
$B \rightarrow \rho$ transitions, i.e. the hard spectator 
interactions. However, such corrections from the dropped diagrams are 
 small as already indicated in \cite{gsw}. 
This point will be discussed in much detail in a 
subsequent work \cite{ah}.
\\

\noindent
\textbf{ACKNOWLEDGEMENTS}\\

The author acknowledges the collaboration with A. Ali and D. London, 
and also greatly appreciates the Organizers for warm 
hospitality during the Conference, and the Alexander von 
Humbodlt Foundation and DESY for their supports.

\end{document}